\documentclass[]{aa}  

\usepackage{graphicx,siunitx}
\usepackage{txfonts}
\usepackage[normalem]{ulem}
\usepackage[]{hyperref}
\hypersetup{backref=true, pagebackref=true,hyperindex=true,breaklinks=true,colorlinks=true,urlcolor=blue,linkcolor=blue,citecolor=blue,pagecolor=red,bookmarks=true,bookmarksopen=true}

\begin{document} 

\title{ALMA chemical survey of disk-outflow sources in Taurus (ALMA-DOT)}
   
   \subtitle{IV. Thioformaldehyde (H$_2$CS) in protoplanetary disks: spatial distributions and binding energies}

   \author{C. Codella
          \inst{1,2}
          \and
           L. Podio\inst{1}
          \and
          A. Garufi\inst{1}
          \and
          J. Perrero\inst{3}
          \and
          P. Ugliengo\inst{3}
          \and
          D. Fedele\inst{1,4}
          \and
          C. Favre\inst{2}
          \and
         E. Bianchi\inst{2}
        \and
         C. Ceccarelli\inst{2}
         \and
         S. Mercimek\inst{1,5}
         \and
         F. Bacciotti\inst{1}
          \and
          K.L.J. Rygl\inst{6}
          \and
          L. Testi\inst{7,8,1}
          }

\institute{
INAF, Osservatorio Astrofisico di Arcetri, Largo E. Fermi 5,
50125 Firenze, Italy
\and
Univ. Grenoble Alpes, CNRS, Institut de
Plan\'etologie et d'Astrophysique de Grenoble (IPAG), 38000 Grenoble, France
\and
Dipartimento di Chimica and Nanostructured Interfaces and Surfaces (NIS) Centre, Universit\`a degli Studi di Torino, via P. Giuria 7, 10125 Torino, Italy
\and
INAF, Osservatorio Astrofisico di Torino, Via Osservatorio 20, I-10025 Pino Torinese, Italy
\and
Universit\`a degli Studi di Firenze, Dipartimento di Fisica e Astronomia, Via G. Sansone 1, 50019 Sesto Fiorentino, Italy
\and
INAF, Istituto di Radioastronomia \& Italian ALMA Regional Centre, via P. Gobetti 101, 40129 Bologna, Italy
\and 
European Southern Observatory, Karl-Schwarzschild-Strasse 2, 85748 Garching bei M{\"u}nchen, Germany
\and 
Excellence Cluster Origins, Boltzmannstrasse 2, 85748 Garching bei M{\"u}nchen, Germany}

\offprints{C. Codella, \email{codella@arcetri.astro.it}}
\date{Received date; accepted date}

\authorrunning{Codella et al.}
\titlerunning{ALMA-DOT: H$_2$CS in protoplanetary disks}

  \abstract
   {Planet formation starts around Sun-like protostars with ages $\leq$ 1 Myr: what is  the chemical compositions in disks?}
   {To trace the radial and vertical spatial distribution of H$_2$CS, a key species of the S-bearing chemistry, in protoplanetary disks. To analyse the observed distributions in light of the H$_2$CS binding energy, in order to discuss the role of thermal desorption in enriching the gas disk component.}
   {In the context of the ALMA chemical survey of Disk-Outflow sources in the Taurus star forming region (ALMA-DOT), we observed five Class I or early Class II sources with the o--H$_2$CS($7_{1,6}-6_{1,5}$) line. ALMA-Band 6 was used, reaching spatial resolutions $\simeq$ 40 au, i.e. Solar System spatial scales.
   We also estimated the binding energy of H$_2$CS using quantum mechanical calculations, for the first time, for an extended, periodic, crystalline ice.}
   {We imaged H$_2$CS emission in two rotating molecular rings in the HL Tau and IRAS04302+2247 disks. The outer radii are $\sim$ 140 au (HL Tau), and 115 au (IRAS 04302+2247). The edge-on geometry of IRAS 04302+2247 allows us to reveal that H$_2$CS emission peaks, at radii of 60--115 au, at $z$ = $\pm$ 50 au from the equatorial plane. Assuming LTE conditions, the column densities are $\sim$ 10$^{14}$ cm$^{-2}$. Upper limits of a few 10$^{13}$ cm$^{-2}$ have been estimated for the H$_2$CS column densities in DG Tau, DG Tau B, and Haro 6-13 disks. For HL Tau, we derive, for the first time, the [H$_2$CS]/[H] abundance in a protoplanetary disk ($\simeq$ 10$^{-14}$). The binding energy of H$_2$CS 
   computed for extended crystalline ice 
   and amorphous ices is 4258 K and
   3000-4600 K, respectively, implying a thermal evaporation where dust temperature is $\geq$ 50--80 K.}
   {H$_2$CS traces the so-called warm molecular layer, a
   region previously sampled using CS, and H$_2$CO. Thioformaldehyde peaks closer to the protostar
  than H$_2$CO and CS, plausibly due to the relatively high-excitation level of observed $7_{1,6}-6_{1,5}$ line (60 K). The H$_2$CS binding energy implies that thermal desorption dominates in thin, au-sized, inner and/or upper disk layers, indicating that the observed H$_2$CS emitting up to radii larger than 100 au is likely injected in the gas  due to non-thermal processes.}


   \maketitle
%

\section{Introduction}

Low-mass star formation is the process starting from a molecular cloud and ending with a Sun-like
star with its own planetary system. According to the classical scenario
\cite[e.g.][and references therein]{Andre2000,Caselli2012}, 
the central object
increases its mass through an accretion disk, while a fast
jet contributes in removing the angular momentum excess. While the star accrete its mass, part of the material in the disk is incorporated to form planets.  As the physical process proceeds, also the chemistry evolves towards a complex gas composition \citep[see][]{Ceccarelli2007,Herbst2009}. 
Several observational programs have been dedicated to the chemical content of {\it protostellar} envelopes/disks using both single-dishes and interferometers at mm-wavelenghts (IRAM 30-m ASAI
\citep{Lefloch2018}; ALMA PILS \citep{Jorgensen2016}; IRAM PdBI CALYPSO
\citep{Belloche2020}; IRAM NOEMA SOLIS \citep{Ceccarelli2017}; ALMA FAUST\footnote{http://faust-alma.riken.jp}, Bianchi et al. 2020).
Nevertheless, the chemical content of the {\it protoplanetary} disks is still poorly known. 
Tiny outer molecular layers (and thus small column densities)
are formed due to thermal or UV-photo/CR-induced desorption \citep[e.g.,][]{Semenov2011,Walsh2014,Walsh2016,Loomis2015,LeGal2019}.
More precisely, the disk portions where thermal desorption is 
expected to rule is in turn determined by the binding energies (BEs) of the species to grains. Different BEs imply different temperatures
of the dust which can inject iced molecules into the gas-phase
\citep[e.g.,][]{Penteado2017}.

A recent blooming of projects focused on 
protoplanetary disks, mainly with ALMA\footnote{Atacama Large Millimeter Array: https://www.almaobservatory.org}, led to the detection of   
several molecules from CO isotopologues to complex species such as t-HCOOH, CH$_3$OH, and CH$_3$CN            
\cite[see][]{Oberg2013,Oberg2015a,Oberg2015b,Oberg2016,Guilloteau2016,Favre2015,Favre2018,Favre2019,Fedele2017,Semenov2018,Podio2019,Podio2020,Podio2020b,Garufi2020b}. The ALMA images allowed the authors to compare  
the gas and dust distribution, starting the first steps of physical-chemical modelling.     
A breakthrough result provided by ALMA, through  
(sub-)mm array observations of young stellar objects, is that planets start to form 
already during the protostellar phases hence before the classical protoplanetary stage
with an age of at least 1 Myr. This is indirectely indicated by the presence of      
rings, gaps, and spirals in disks with ages less than 1 Myr 
\citep[see e.g.,][]{Sheehan2017,Fedele2018,Andrews2018}. These substructures have been observed also using molecules, driving studies to sample the molecular components of disks. This is the goal of the ALMA-DOT project (ALMA chemical survey of Disk-Outflow sources in Taurus), which targets Class I or early Class II disks to obtain their chemical characterization.

\subsection{The S-bearing molecules}

Chemistry of S-bearing species is not well understood.
In dense gas, which is involved in the star forming process,
sulphur is severely
depleted \citep[e.g.][]{Wakelam2004,Phuong2018,Tieftrunk1994,
Laas2019,vanthoff2020}, by at least two orders of magnitude with respect to the Solar System value [S]/[H] = 1.8 $\times$
10$^{-5}$ \citep{Anders1989}.
The main S-carrier species on dust grains, however, are still unknown.
For years, H$_2$S has been postulated to be the solution, but so far
it has been never directly detected on interstellar ices \citep{Boogert2015}.
Alternative solutions have been proposed in light of studies focused on protostellar shocks,  
where dust is sputtered: S, OCS, or H$_2$CS \citep[e.g.][]{Wakelam2004,Codella2005,Podio2014,Holdship2016}, but, again,       
no detection on ices has been reported \citep{Boogert2015}. 
What about the inventory of S-molecules in protoplanetary disks?
Only four species have been detected, often by single-dish:  CS, SO, H$_2$S, and H$_2$CS.
More specifically, multiline CS emission has been observed towards  $\geq$ 10 disks
\citep[e.g.][]{Dutrey1997,Dutrey2017,Fuente2010,Guilloteau2013,Guilloteau2016,Teague2018,Phuong2018,Semenov2018,LeGal2019,Garufi2020b,Podio2020}.
SO emission has been detected in fewer disks, part of them still
associated with accretion, as e.g. TMC-1A  \citep[e.g.][]{Fuente2010,Guilloteau2013,Guilloteau2016,Sakai2016,Pacheco2016,Teague2018,Booth2018}. On the other hand, only very recently 
H$_2$S and H$_2$CS have been detected and imaged towards a couple of disks.
Namely, H$_2$S has been imaged towards GG Tau A  \citep{Phuong2018}, while H$_2$CS has been observed with ALMA towards  MWC480, and, tentatively, LkCa 15 \citep{LeGal2019,Loomis2020}. Further observations of the S-species supposed to be the main S-carrier on dust grains are required. 

In this context, the goal of the present project is twofold: (1) to map the H$_2$CS (thioformaldehyde) spatial distribution in protoplanetary disks in an intermediate phase between Class I and Class II, and (2) to derive the binding energies of H$_2$CS using quantum mechanical calculations for an extended crystalline ice. We focus on H$_2$CS in the effort to identify S-bearing species able to trace protoplanetary disks.
The article is organized as follows: 
we first present the observations and data reduction process (Sect. 2), then we report the observational results (Sect. 3), and present
the method and the assumptions used to derive BEs (Sect. 4).
In Sect. 5, the results will be discussed by comparing 
the H$_2$CS maps with (i) those of other 
molecular species as well as with (ii) the spatial distributions expected in case the thermal desorption process (driven by the BE values) is
the main mechanism leading to H$_2$CS in the gas-phase.
Finally, Sect. 6 summarises our work.

\section{Observations: sample and data reduction}

The sample consists of four Class I and one early Class II 
\citep[e.g.][]{Andre2000} well known sources
\citep{Guilloteau2013,Guilloteau2014}: DG Tau, DG Tau B, HL Tau, 
IRAS 04302+2247, and Haro 6--13 (aka V 806 Tau).
The sources are observed in the context of the ALMA-DOT project (ALMA chemical survey of Disk-Outflow sources in the Taurus star forming region, \citealt{Podio2019,Podio2020,Garufi2020b}), which targets sources: (i) still embedded in a dense envelope, (ii) driving an atomic jet, and a molecular outflow.
The sources are chemically rich as revealed by IRAM-30m observations detecting CO isotopologues, H$_2$CO, and CN (plus SO for all but Haro 6--13) \citep{Guilloteau2013}.

This work is based on ALMA Cycle 4 observations of DG Tau and DG Tau B and Cycle 6 observations of HL Tau, IRAS 04302+2247, and Haro 6-13 (projects 2016.1.00846.S and 2018.1.01037.S, PI: L. Podio).
The DG Tau and DG Tau B observations were described by \citet{Podio2019}, and \citet{Garufi2020b}, respectively.
All Band 6 observations were taken in an extended array configuration with baselines ranging from 15 m to 1.4 km or from 17 m to 3.7 km. The frequency interval 
covered by the continuum spectral window 
included the o--H$_2$CS($7_{1,6}-6_{1,5}$) line emitting at 244048.5 MHz, characterised\footnote{The spectral parameters \citep{Maeda2008} are taken from the Cologne Database for Molecular Spectroscopy \citep{Mueller2005}.} by $E_{\rm up}$ = 60 K, and $S\mu^2$ = 56 D$^2$. A standard data reduction was performed with CASA pipeline version 4.7.2. Self-calibration was performed on the continuum emission and then applied to the continuum-subtracted line datacube. The r.m.s. for continuum images are about 68 $\mu$Jy (HL Tau) and 40 $\mu$Jy (IRAS 04302+2247).
The line spectral cube was produced through \textsc{tclean}. We used robust weighting in order to maximise the spatial resolution, and set a channel width of 1.2 km s$^{-1}$. The r.m.s. per each channel is about 0.8 mJy beam$^{-1}$.
The synthesized beam (HPBW) is $\sim$ 
0$\farcs$3 $\times$ 0$\farcs$3. 

\section{Observational results}

\subsection{Continuum and H$_2$CS spatial distributions}

\begin{figure*}
\centerline{\includegraphics[angle=0,width=14cm]{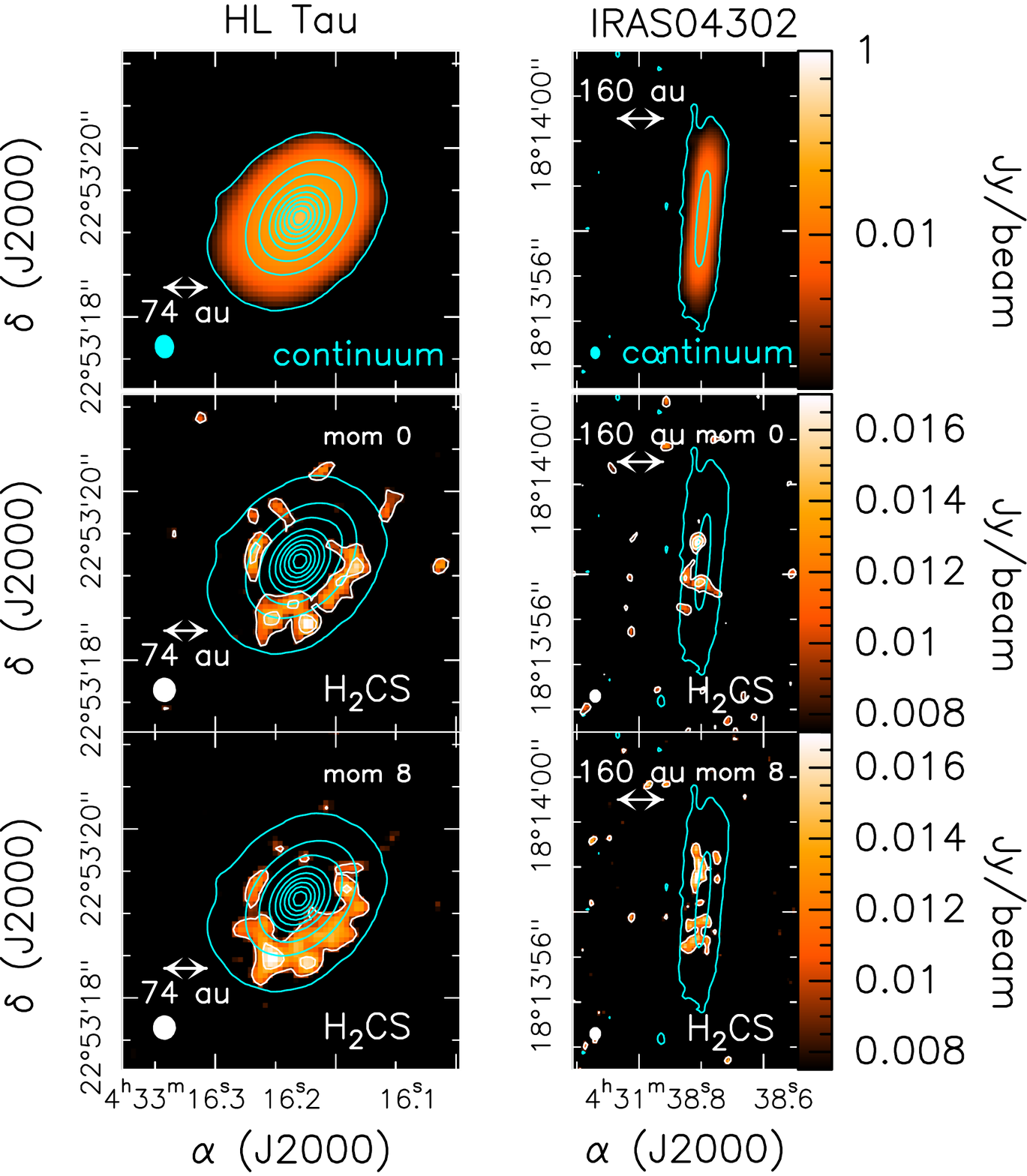}}
\caption{{\it Upper panels:} Map (cyan contours and colour scale) of the 1.3mm dust continuum distribution
for HL Tau (Left) and IRAS 04302+2247 (Right) disks. First contours and steps are
3$\sigma$ (200 $\mu$Jy beam$^{-1}$, HL Tau; 115 $\mu$Jy beam$^{-1}$, IRAS 04302+2247), and 200$\sigma$, respectively.
The ellipse in the bottom left corner shows the ALMA synthesized beam (HPBW): 
0$\farcs$28 $\times$ 0$\farcs$25 (PA = --7$\degr$), for HL Tau, and
0$\farcs$28 $\times$ 0$\farcs$22 (PA = --3$\degr$), for IRAS 04302+2247.
{\it Middle panels:} Spatial distribution (moment 0) maps (white contours and colour scale) 
of the o--H$_2$CS(7$_{\rm 1,6}$--6$_{\rm 1,5}$) line based on the velocity integrated emission (3--11 km s$^{-1}$,
HL Tau, 2--10 km s$^{-1}$, IRAS 04302+2247), overlaid on the continuum maps (cyan contours).
First contours and steps are
5$\sigma$ (7.5 mJy beam$^{-1}$ km s$^{-1}$, HL Tau, and 12.5 mJy beam$^{-1}$ km s$^{-1}$, IRAS04302), and 3$\sigma$, respectively.
The ellipse shows the synthesized beam (HPBW):
0$\farcs$29 $\times$ 0$\farcs$27 (PA = 2$\degr$), for HL Tau, and
0$\farcs$30 $\times$ 0$\farcs$26 (PA = 2$\degr$), for IRAS 04302+2247.
{\it Lower panels:} Moment 8 maps (white contours and colour scale) 
of the H$_2$CS(7$_{\rm 1,6}$--6$_{\rm 1,5}$) based on the same velocity integrated emission used for the moment 0 maps, overlaid on the continuum maps (cyan contours).
First contours and steps are
5$\sigma$ (3 mJy beam$^{-1}$ km s$^{-1}$), and 3$\sigma$, respectively.}
\label{distributions}
\end{figure*} 

Out of the 5 observed targets, we detected o--H$_2$CS($7_{1,6}-6_{1,5}$) emission
towards 2 disks, namely HL Tau and IRAS 04302+2247 (herafter IRAS04302). 
For DG Tau, DG Tau B, and Haro 6--13 upper limits on
H$_2$CS emission will be reported in the next sections.
As a reference to analyse molecular emission, Figure 1 (upper panels) reports the dust continuum emission. The well known HL Tau disk \citep[$d$ = 147 pc;][]{Galli2018}, with
an inclination angle $i$ of 47$\degr$ \citep[e.g][]{ALMA2015,Carrasco2019},
is well traced, showing a radius of $\sim$ 150 au. 
The dust continuum at 227 GHz peaks (118 mJy beam$^{-1}$) at 
$\alpha({\rm J2000})$ = 04$^h$ 31$^m$ 38$\fs$44,
$\delta({\rm J2000})$ = +18$\degr$ 13$\arcmin$ 57$\farcs$65. 
On the other hand, the IRAS04302 disk \citet{Guilloteau2013,Podio2020}, located at $d$ = 161 pc \citep{Galli2019}, is more extended ($\sim$ 350 au), and is associated with an almost edge-on geometry, $i$ $\sim$ 90$\degr$ \citep{Wolf2003}. 
The coordinates of the continuum peak, 133 mJy beam$^{-1}$, are
$\alpha({\rm J2000})$ = 04$^h$ 33$^m$ 16$\fs$47,
$\delta({\rm J2000})$ = +22$\degr$ 53$\arcmin$ 20$\farcs$36.

Figure 1 (Middle) reports also the intensity integrated maps 
(moment 0) of the o--H$_2$CS($7_{1,6}-6_{1,5}$) emission
as observed towards HL Tau and IRAS04302,
thus providing for the first time the 
thioformaldeyde radial (HL Tau) and vertical
(IRAS04302) distributions.
All the channels showing emission of at least 3$\sigma$ (see Sect. 2) have been used,
namely: 3--11 km s$^{-1}$ (HL Tau) and 2--10 km s$^{-1}$ (IRAS04302).
The Signal-to-Noise (S/N) 
of the velocity integrated emission 
is, in both sources, larger than 8.
For HL Tau, the image clearly shows an H$_2$CS ring around a central dip. The outer radius is about 140 pc, while
the dip is confined in the inner $\sim$ 35 au.
For IRAS04302, the picture as provided by the mom 0 map is less clear. Surely, the H$_2$CS emission is confined in
the inner 0$\farcs$7, $\sim$ 115 au. In Sect. 3, we will show how kinematics will lead us to
infer the structure of the emitting region.

Finally, Figure 1 (Bottom panels) reports, in colour scale, the 
HL Tau and IRAS04302 H$_2$CS($7_{1,6}-6_{1,5}$) spatial distributions of the peak intensity
as derived using the moment 8 method\footnote{https://casa.nrao.edu/Release3.4.0/docs/UserMan/UserManse41.html} to enlight the images.
The moment 8 CASA algorithm has been used by collapsing the intensity axis of the ALMA datacube to one pixel and setting the value of that pixel (for R.A. and Dec.) to the maximum value of the spectrum. The moment 8 images definitely confirm 
what found with the moment 0 maps.
In Sect. 5.2, the spatial distribution will be compared with those of other molecular species.

\begin{figure*}
\centerline{\includegraphics[angle=0,width=16cm]{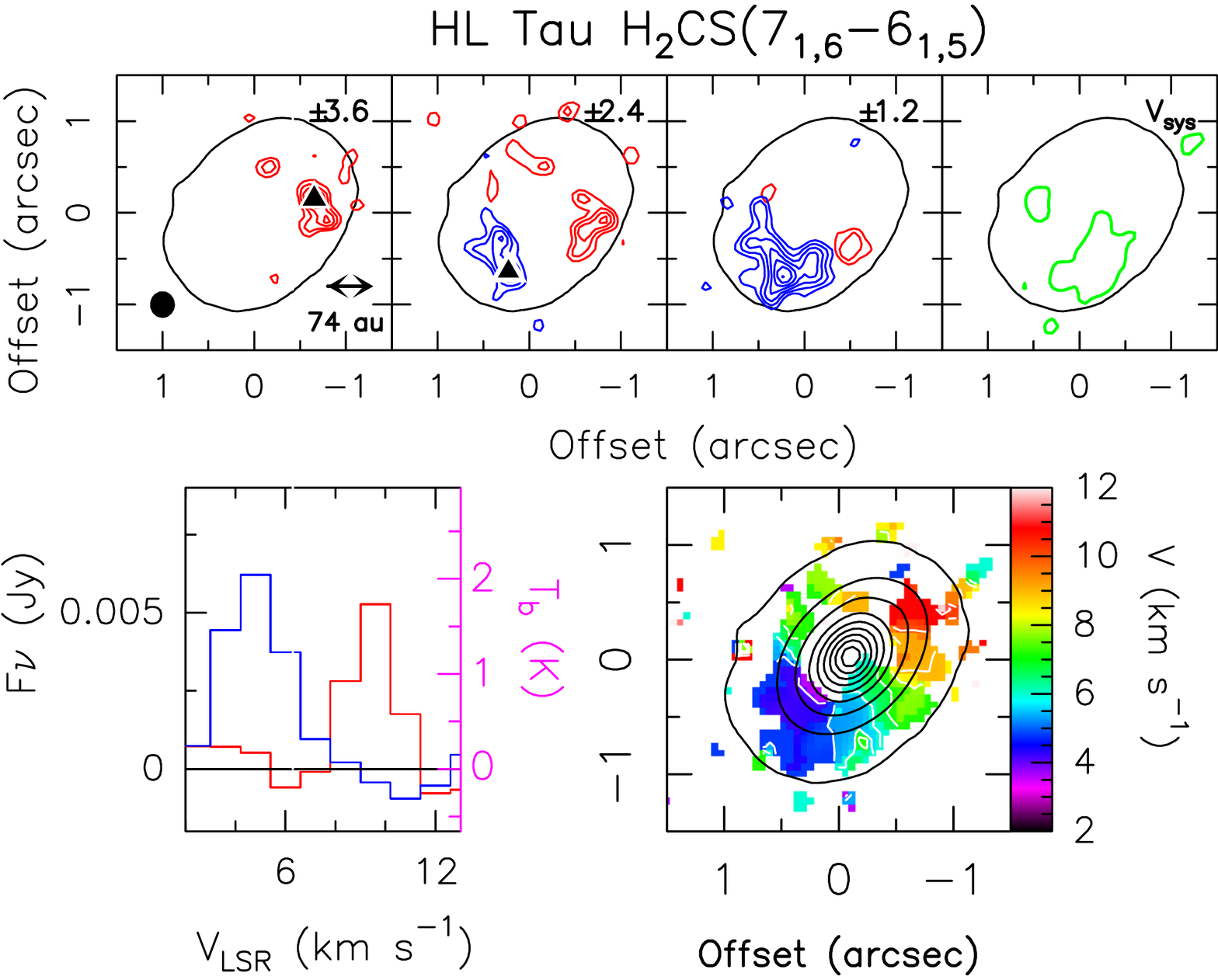}}
\caption{{\it Upper:} Channel maps of the o--H$_2$CS(7${\rm 1,6}$--6${\rm 1,5}$)
blue- and red-shifted emission in the HL Tau disk. Each panel shows the
emission integrated over a velocity interval of 1.2 km s$^{-1}$ shifted with respect to the
systemic velocity ($\sim$ +7 km s$^{-1}$, green) by the value given in the upper-right corner.
We report, in black, the 3$\sigma$ contour of the continuum emission (Figs. 1, A.1).
The black triangles indicate where spectra have been extracted.
The ellipse shows the synthesized beam (HPBW): 0$\farcs$29 $\times$ 0$\farcs$27 (PA = 2$\degr$).
First contours and steps correspond to 3$\sigma$ (2.1 mJy beam$^{-1}$) and 1$\sigma$,
respectively. Offsets are derived with respect to  the continuum peak (Sect. A). {\it Bottom Left:} o--H$_2$CS(7${\rm 1,6}$--6${\rm 1,5}$) spectra in 
flux and brightness temperature scales ($T_{\rm b}$/$F_{\rm \nu}$ = 328.593) extracted in the positions marked with a red
or blue triangle in the Upper panels. 
{\it Bottom Right:} First-moment map in colour scale.}
\label{HLTauchannel}
\end{figure*}

\begin{figure*}
\centerline{\includegraphics[angle=0,width=12cm]{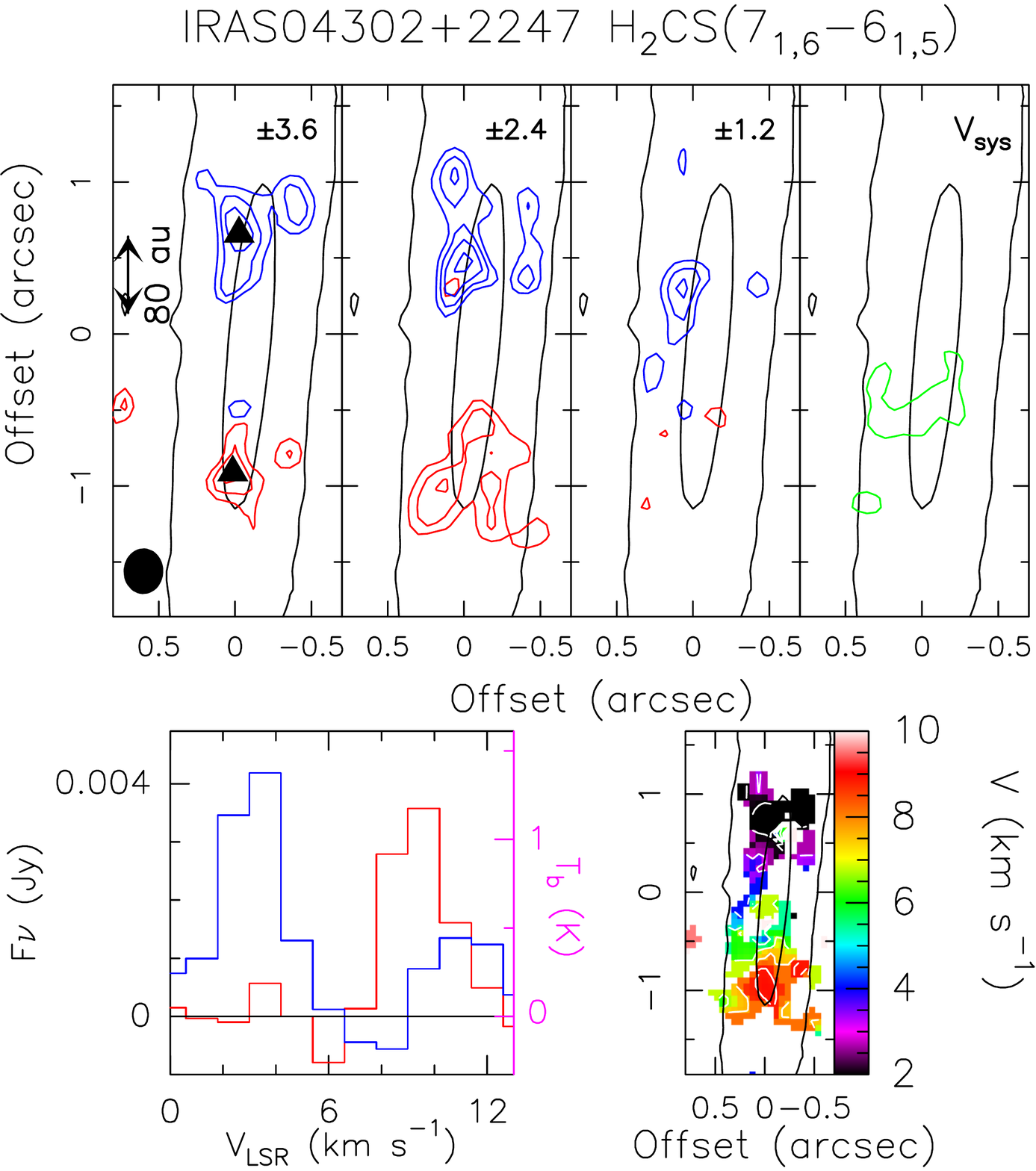}}
\caption{{\it Upper:} Channel maps of the o--H$_2$CS(7$_{\rm 1,6}$--6$_{\rm 1,5}$)
blue- and red-shifted emission in the IRAS 04302+2247 disk. Each panel shows the
emission integrated over a velocity interval of 1.2 km s$^{-1}$ shifted with respect to the
systemic velocity 
\citep[$\sim + 5.6$ km s$^{-1}$][green]{Podio2020}
by the value given in the upper-right corner.
We report,in black, the 3$\sigma$ contour of the continuum emission (Figs. 1, A.1).
The black triangles indicate where spectra have been extracted.
The ellipse shows the synthesized beam (HPBW):
0$\farcs$30 $\times$ 0$\farcs$26 (PA = 2$\degr$). First contours and steps correspond to 3$\sigma$ (2.1 mJy beam$^{-1}$) and 1$\sigma$,
respectively. Offsets are derived with respect to the continuum peak (Sect. A). {\it Bottom Left:} o--H$_2$CS(7$_{\rm 1,6}$--6$_{\rm 1,5}$) spectra in
flux and brightness temperature scales ($T_{\rm b}$/$F_{\rm \nu}$ = 328.593) extracted in the positions marked with a red
or blue triangle in the Upper panels. 
{\it Bottom Right:} First-moment map in colour scale.}
\label{IRAS04channel}
\end{figure*}

\subsection{H$_2$CS kinematics}

Figures 2 and 3 summarise the kinematics of HL Tau and IRAS04302 provided by: (i) channel maps,
(ii) intensity weighted mean velocity (moment 1) maps, and (iii) H$_2$CS
spectra as extracted towards the brightest positions. The H$_2$CS in HL Tau is clearly rotating with the blue- and red-shifted sides located towards SE
and NW, respectively. The systemic velocity is $\sim$ +7.0 km s$^{-1}$ \citep[in agreement with][]{ALMA2015,Wu2018}. The same rotating gradient has been observed using HCO$^+$ \citep{ALMA2015,Yen2019}, C$^{18}$O \citep{Wu2018}, and
$^{13}$C$^{17}$O \citep{Booth2020}. 
Also the emission towards IRAS04302 disk shows a rotation pattern.
Its almost edge-on
orientation allows us to clearly disentangle the red- (southern) and blue-shifted (northern) lobes.
The systemic velocity estimated from the velocity distribution of the bright H$_2$CO line is $+5.6$ km s$^{-1}$ \citep{Podio2020},
in good agreement with IRAM 30-m
CN, H$_2$CO, CS, and C$^{17}$O spectra of
\cite{Guilloteau2013,Guilloteau2016}.
The rotation pattern
has been imaged also by \citet{Podio2020},
who found, using CO, H$_2$CO, and CS emission, a molecular emission which is vertically stratified (see Sect. 4 for the comparison with H$_2$CS).

The H$_2$CS channel maps show that the emission shifted by less than 2 km s$^{-1}$ is
emitted from the inner 0$\farcs$5 region, whereas
at larger velocities we detect emission up to 1$\farcs$2 from the protostar. This suggests that larger projected velocities 
(in particular the blue-shifted velocity) 
have a larger projected positional offset from the protostar. 
In other words, the present dataset suggests 
that V $\propto$ R. 
This is the standard signature of a rotating ring with an inner dip,
and not a filled disk, where 
we would have an opposite trend, with V $\propto$ R$^{-2}$. Very recently, \citet{Oya2020} reported the detection of H$_2$CS in the IRAS16293 A protostellar disk \citep[see also][]{vanthoff2020}. In addition, the present findings well resamble
what found in the archetypical protostellar
disk HH212, also close to be edge-on as
IRAS04302 \citep[e.g.][and references therein]{lee2019}. Also in this case, a chemical enrichment in the gas phase associated
with rotating rings is revealed by the
V $\propto$ R kinematical feature
\citep[see e.g. the recent review by][]{Codella2019}.

\subsection{H$_2$CS column densities and abundances}

Assuming (i) Local Thermodynamic
Equilibrium (LTE), and
(ii) optically thin emission, the column densities of the H$_2$CS are derived. 
The first assumption is well justified as the H$_2$ gas density in the molecular layers is larger than 10$^7$ cm$^{-3}$ \citep[see e.g.][]{Walsh2014,LeGal2019}, i.e. well above the critical density of the considered 
o-H$_2$CS(7$_{\rm 1,6}$--6$_{\rm 1,5}$) line ($n_{\rm cr}$ $\sim$ a few 10$^4 $ cm$^{-3}$ in the 20--150 K range\footnote{Using scaled H$_2$CO collisional rates from \citet{Wiesenfeld2013}, see \citet{Schoier2005}.}.
The second assumption is justified as models indicate an abundance (with respect to H) of H$_2$CS lower than 10$^{-10}$ \citep{LeGal2019,Loomis2020}.

In HL Tau the observed emitting area is 1.35 arcsec$^2$ (from the moment 0 map) the flux is 148 mJy km s$^{-1}$ (2.6 K km s$^{-1}$).  \cite{LeGal2019} estimated 
an H$_2$CS rotational temperature between 20 K and 80 K in the MWC 480 disk. We conservatively adopted a temperature  between 20 K and 150 K. An ortho-to-para ratio of 3, i.e. the statistical value, has been assumed \citep[as done by][]{LeGal2019}. 
The total H$_2$CS column density (averaged on the emitting area)  $N_{\rm  H_2CS}$ turns out to be 0.9--1.4 $\times$ 10$^{14}$ cm$^{-2}$.
For IRAS04302 the emitting area is smaller with respect to HL Tau, 0.72 arcsec$^2$ and the flux is $\simeq$ 78 mJy km s$^{-1}$ (2.5 K km s$^{-1}$).  The column density $N_{\rm  H_2CS}$ is, similarly to HL Tau, $\simeq$ 10$^{14}$ cm$^{-2}$.
These values can be also considered as 
an a-posteriori check that 
the observed H$_2$CS emission is optically thin.
Indeed, by assuming kinetic temperatures larger than 20 K and densities larger than 10$^7$ cm$^{-3}$, the Large Velocity Gradient approach confirms that column densities around 10$^{14}$ cm$^{-2}$
implies for the o-H$_2$CS(7$_{\rm 1,6}$--6$_{\rm 1,5}$)
line an opacity of $\leq$ 0.1.
These column density estimates are larger by a factor 30 with respect to the ones  measured by \cite{LeGal2019} and \cite{Loomis2020} towards the massive disk MWC 480 (3 $\times$ 10$^{12}$ cm$^{-2}$). However, as reported
by the authors, their spatial resolution is not high enough to
evaluate the presence of a central dip, which could cause an underestimate the H$_2$CS column density. 

For HL Tau a further step can be done given that  \cite{Booth2020} published an ALMA map of the $^{13}$C$^{17}$O(3--2) emission at 
a similar angular scale as our H$_2$CS maps.
The $^{13}$C$^{17}$O spatial distributions well agrees with that of H$_2$CS.
 The emission is considered optically thin, given that its abundance ratio with respect to $^{12}$C$^{16}$O is 8.3 $\times$ 10$^{-6}$
 \citep{Milam2005}. Taking the $^{13}$C$^{17}$O flux density of 10-20 mJy beam$^{-1}$ km s$^{-1}$, and an excitation temperature, as for H$_2$CS, in
 the 20-150 K range, we derive $N_{\rm  CO}$ $\simeq$ 5--10 $\times$ 10$^{23}$ cm$^{-2}$. Using [CO]/[H] = 5 $\times$ 10$^{-5}$, the H$_2$CS abundance (with respect to H) results to be $X(H_2CS)$ $\simeq$ 10$^{-14}$.
The comparison with the H$_2$CS abundances predicted
by disk chemistry is challenging, given this 
is dramatically depending on
the initial composition of the S-species, which
can vary by orders of magnitude
\citep[see e.g][]{Fedele2020}.
Vice versa, the present measured H$_2$CS abundances will be hopefully used in future chemical models to 
constrain their initial conditions.

Finally, the upper limits on the H$_2$CS column densities in DG Tau, DG Tau B, and Haro 6-13, derived
taking the 5$\sigma$ level of the moment 0 maps, and using all the assumptions as above, are:
$N_{\rm  H_2CS}$ $\leq$ 2 $\times$  10$^{13}$ cm$^{-2}$ (DG Tau B), and $N_{\rm  H_2CS}$ $\leq$ 4 $\times$  10$^{13}$ cm$^{-2}$ (DG Tau and Haro 6-13).


\section{Binding energies: ice modelling and computational methods}

We computed the H$_2$CS and CS BEs, adopting for the bulk ice a proton-ordered (P-ice) crystalline model \citep{Casassa1997}.
The ice surface where the adsorption takes place was simulated by a finite slab model of the (010) surface cut out from the bulk P-ice crystal, as recently proposed by  \cite{Ferrero2020} to predict the BEs of a set of 21 molecules using the periodic \textit{ab initio} program \textsc{CRYSTAL17} \citep{Dovesi2018}. CRYSTAL17 adopts localised (Gaussian) basis functions which allow to simulate the surfaces as a true 2D systems, without including fake replicas of the slab separated by artificial voids. 
The surface slab model is thick enough (number of water layers) to ensure a converged surface energy (energy penalty to cut the surface from the bulk ice). The choice of a crystalline ice model is against the overwhelming evidence that ice in the interstellar environment is of amorphous nature (AWS). There are two main reasons that hinder the adoption of AWS phase at the modeling level: i) the experimental atomistic structure of the AWS is unknown; ii) AWS model should be based on very large unit cell to mimic the disorder nature of the ice. Point i) means that the model cannot be derived from experimental evidence and therefore any model is somehow arbitrary. Point ii) implies that very expensive calculations are needed to cope with the large unit cells. 
In the following, we proposed a simplified strategy to derive the values of the BEs for the H$_2$CS and CS molecules as adsorbed on AWS, without actually run the needed expensive calculations, but taking profit of the results by \cite{Ferrero2020} on the analogous CO and H$_2$CO molecules, in which both crystalline and AWS models were studied.

\begin{figure}
\centerline{\includegraphics[angle=0,width=8cm]{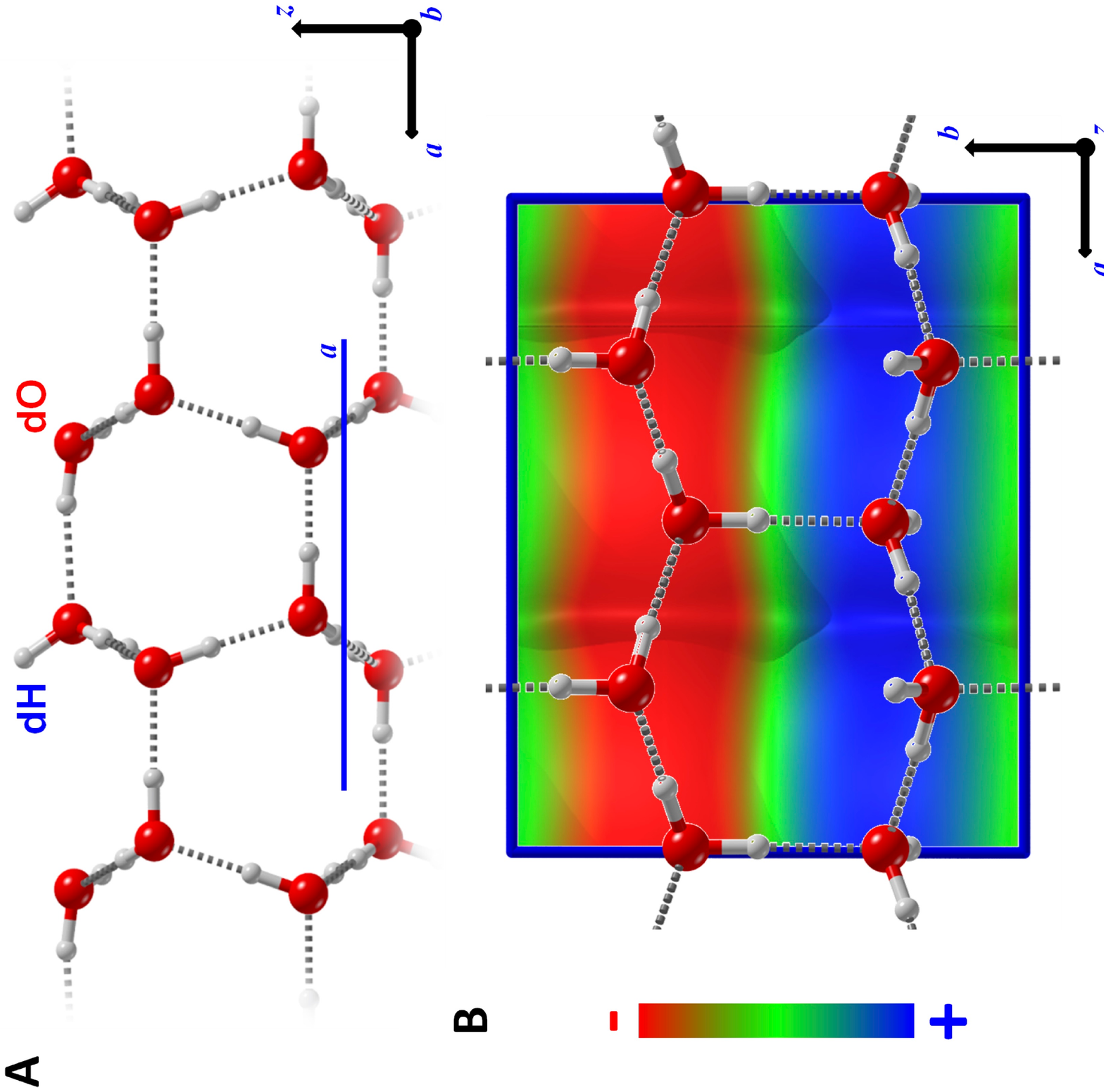}}
\caption{HF-3c optimized P-ice slab model. The adsorption crystallographic plane is the Miller (010) one. 
{\it A Panel:} Side view along the \textit{b} lattice vector. {\it B Panel:} Top view of the 2 $\times$ 1 supercell ($|a| = \SI{9.065}{\angstrom}$ and $|b| = \SI{7.154}{\angstrom}$) along with its ESP map. Colour code: +0.02 
atomic unit (blue, positive), 0.00 atomic unit (green, neutral) and --0.02 atomic unit (red, negative).}
\label{ice-esp}
\end{figure}

First, we set up the starting locations of H$_2$CS and CS molecules at the ice surface using the optimized positions of the analogous CO and H$_2$CO molecules after \cite{Ferrero2020} and then fully optimizing the structures.
We choose the HF-3c method, which combines the Hartree-Fock Hamiltonian with the minimal basis set MINI-1 \citep{Tatewaki1980}
supplemented by three \emph{a posteriori} corrections for: (i) the basis set superposition error (BSSE), arising when localized Gaussian functions are used to expand the basis set \citep{Jansen1969,Liu1973};
(ii) the dispersive interactions; (iii) short-ranged deficiencies due to the adopted minimal basis set \citep{Sure2013}. 
The  resulting optimized unit cell of the ice (010) surface  is shown in Figure \ref{ice-esp}, together with the mapping of the electrostatic potential (ESP) computed at HF-3c level of theory. 
The ESP reveals the oxygen rich (red color, ESP $<$ 0) and proton rich (blue color, ESP $>$ 0) regions, acting, respectively, as a H-bond acceptor/donor with respect to specific adsorbates.

In order to get accurate BEs, the HF-3c structures were used to compute single point energy evaluation at B3LYP-D3(BJ) level of theory \citep{Becke1993}. Dispersion interaction is included through the D3 correction, with the Becke-Johnson damping scheme \citep{Grimme2010}. The Ahlrichs' triple-zeta quality VTZ basis set, supplemented with a double set of polarization functions \citep{Schafer1992} was adopted to define the final B3LYP-D3(BJ)/A-VTZ*//HF-3c model chemistry. BEs were also corrected for BSSE. 

For H$_2$CS, the most stable structure shows a symmetric involvement of the CH$_2$S molecule with the ice surface, in which the C-H bonds act as weak H-donors toward the surface oxygen atoms, while the S atom as a H-bond acceptor at the dangling surface OH groups. For CS, the adsorption occurs through the C-end with a very long H-bond with the dangling ice OH group. Any attempt to engage the CS molecule through the S end of the molecule evolved spontaneously to the C-end one. 
Obviously, the relatively simple structure of the crystalline P-ice does not allow to explore other configurations than the described ones, at variance with the results for an AWS model which would be expected to provide a many-fold of adsorption sites of different strength. Furthermore, due to the long range order of the water molecules in the crystalline ice, the H-bond interactions will cooperate to enhance the H-bond donor/acceptor character of the surface water molecules, giving very high/low values of the ESP (see Figure \ref{ice-esp}). This means that the BEs resulting from the crystalline model should be taken as an upper limit, as shown by \cite{Ferrero2020}.
As anticipated, to remedy the deficiencies of the crystalline ice model we resort to data from the work by \cite{Ferrero2020}  for the analog CO and H$_2$CO molecules (with respect to CS and H$_2$CS treated here) adsorbed on the AWS model. As expected, they found a rather complex adsorption scenario at AWS, when compared to that occurring at the P-ice surface. For CO, five different adsorption sites have been characterized; for H$_2$CO, up to eight different adsorption sites were predicted. Obviously, each of these adsorption sites provides different values of the BE and in the limit of very large AWS models, the BE values will obey to a certain distribution. 
It is clear that a similar painstacking study should be carried out also for CS and H$_2$CS. Instead, as anticipated, we resort to a simplified scheme to guess the BE values for the AWS, without actually running any calculation. 

First, from the data by \cite{Ferrero2020}, we worked out two scaling factors (0.859 and 0.775, respectively) connecting the BE of the CO and H$_2$CO analogs  computed for the P-ice to their corresponding averaged BE values for the AWS model. Then by assuming the same scaling factors also for the present CS and H$_2$CS cases, we scaled the BEs for the crystalline ice to estimate their averaged values on the AWS model.
In agreement with the CO and H$_2$CO cases, also for CS and H$_2$CS the BE-AWS are smaller than those for the crystalline ice. 
Table \ref{tbe} reports, for both H$_2$CS and CS, the derived BEs on crystalline (H$_2$CS: 4258 K; CS: 3861 K) and AWS 
(H$_2$CS: 3000-4600 K; CS: 2700-4000 K) ice.

\section{Discussion}

\subsection{Comparison with previous BEs measurements}

Interstellar molecules are formed either through gas-phase reactions or directly on grain surfaces.
Regardless of the formation route, gaseous molecules freeze out into the grain mantles in
timescales that depend on the density and temperature of the gas and dust as well as the molecule BE. Thus, in cold and dense regions, as
in the outer disk regions and in the layers close to the disk midplane, icy mantles envelope the dust grains. The frozen molecules can then be injected into the gas-phase via three processes \citep[see e.g.][]{Walsh2014}: (i) thermal desorption, (ii) UV-photo and/or CR-induced desorption, and (iii) reactive desorption.
The first process occurs inside the so called snow lines, i.e. the location where the dust temperature is high enough to allow the species to sublimate.


Regarding H$_2$CO, its BE has been experimentally measured on amorphous water surface (AWS) through thermal desorption process, e.g. by \cite{Noble2012} to be 3260$\pm$60 K, and 
estimated via theoretical computations on AWS and crystalline ice models by \citet{Ferrero2020}, who found {zero point corrected BEs (see Sect. 4) in the 3071-6194 K range, for the AWS case, therefore bracketing the experimental value. The BE for the crystalline ice falls at 5187 K, in the higher regime of the AWS range.

\begin{table}
   \caption {H$_2$CS and CS BEs (Kelvin) derived for the crystalline (CRY) and amorphous (AWS) ice models. See the text for the procedure adopted to arrive to the BE-AWS values}
    \centering
\begin{tabular}[h]{lcccc} 
\hline
  & BE-CRY & BE-AWS & Das$^a$ &   
 Wakelam$^b$ \\
 \hline
H$_2$CS & 4258  & 3000--4600 & 3110    &  4400   \\
CS & 3861    &   2700--4000   &  2217     &    3200   \\
\hline
\label{tbe}
\end{tabular}

$^a$ \cite{Das2018}; 
$^b$ \citet{Wakelam2017}.
\end{table}

Following the described procedure, we found for H$_2$CS BE values of  4258 K and  3000–4600 K for the crystalline and AWS ice models, respectively.
For CS, we have slightly lower values:  3861 K (crystalline) and 2700--4000 K (AWS).
Table \ref{tbe} compares our results with the available literature data, all coming from computer simulations. The BE computed by \citet{Das2018}, who adopted a tetramer of water molecules to simulate the ice, are in the low end of our computed BEs range, whereas the BE computed by \citet{Wakelam2017}, based on one single water and corrected with an empirical factor, lies on the high end of our range.

Based on our new computations of Sect. 4 and using a BE ranging from 3000 to 4600 K, H$_2$CS is expected to thermally sublimate in regions of the disk where the dust temperature 
exceeds $\sim$50 to $\sim$80 K, respectively.  We emphasise that the dispersion in our computed BEs
reflects the different possible sites of adsorption of H$_2$CS and it is, therefore, physical and not due to a computational uncertainty \citep[see the discussion in][]{Ferrero2020}. Although
we cannot a priori say how many sites with each different BE are populated, we can conservatively assume that H$_2$CS molecules should remain
frozen in regions of the disk with dust temperatures lower than 50--80 K. The
fact that H$_2$CS emission up to outer radii of $\sim$ 140 au (HL Tau) and $\sim$ 115 au 
(IRAS 04302) suggests that 
non thermal processes are likely responsible for the presence of H$_2$CS in the gas in those 
disk regions.


\subsection{Comparison between H$_2$CS, CS, and H$_2$CO}

\begin{figure}
\centerline{\includegraphics[angle=0,width=8cm]{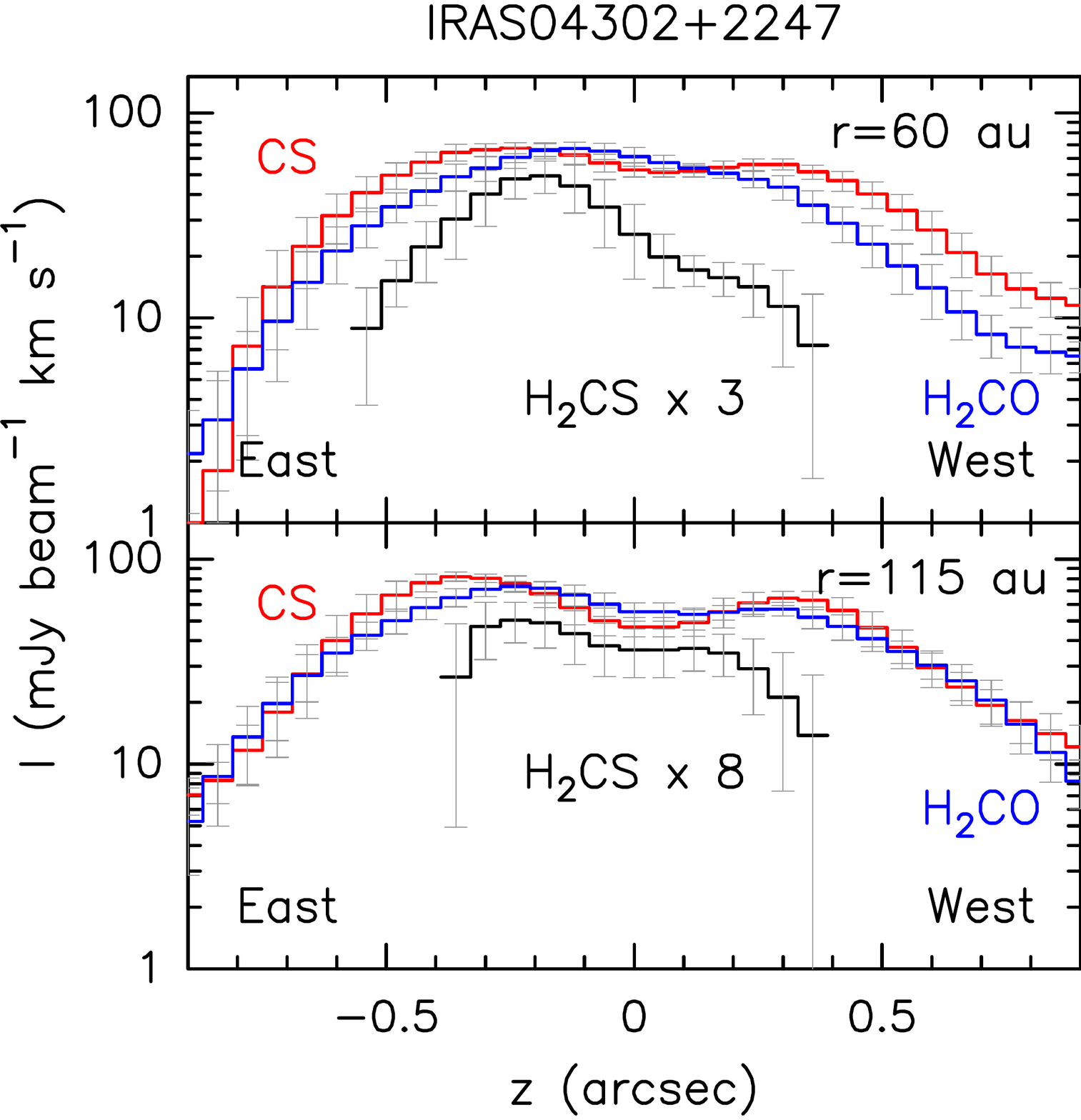}}
\caption{Vertical intensity profile $z$ of H$_2$CS (black) compared with those of CS(5--4) (red) and  o-H$_2$CO($3_{1,2}-2_{1,1}$) 
(blue), from \citet{Podio2020}. Only  fluxes  above  3$\sigma$ confidence  are shown.
The H$_2$CS intensity has been scaled in order to better
compare its profile with those of o-H$_2$CO and CS.
The inner 1$\farcs$8 region is shown, with the positive  (negative) values sampling the eastern (western) side. The angular resolution is 0$\farcs$25 (40 au). The profiles obtained at $\sim$ 60 au (Upper panel) 
and 115 au (Lower panel) from the protostar.}
\label{IRAS04302_radial}
\end{figure}

The formation routes of H$_2$CS have been
recently summarised by \cite{LeGal2019}, who used 
the the gas-grain chemical model {\it Nautilus} \citep{Wakelam2016} supported by
the gas chemical dataset {\it KIDA} \citep{Wakelam2015}, and by \citet{Fedele2020}, who
adopted the thermo-chemical model {\it DALI} \citep{Bruderer2012},
coupled with the chemical network {\it UMIST} \citep{Woodall2007}.
Generally, H$_2$CS is mainly (99\%) formed in gas-phase via neutral–neutral reaction of atomic S with CH$_3$.
On the other hand, CS in disks is also thought to be mainly formed in gas-phase, through reactions starting with small hydrocarbons interacting with S$^+$ (in upper disk layers) or S (in inner slabs).
Finally, H$_2$CO can be formed on grains due to hydrogenation processes as well and in gas-phase by oxygen reacting with CH$_3$, in a analogous way of
S + CH$_3$ $\to$ H$_2$CS \citep[see e.g.][and references therein]{Fedele2020}. 

\begin{figure}
\centerline{\includegraphics[angle=0,width=7.5cm]{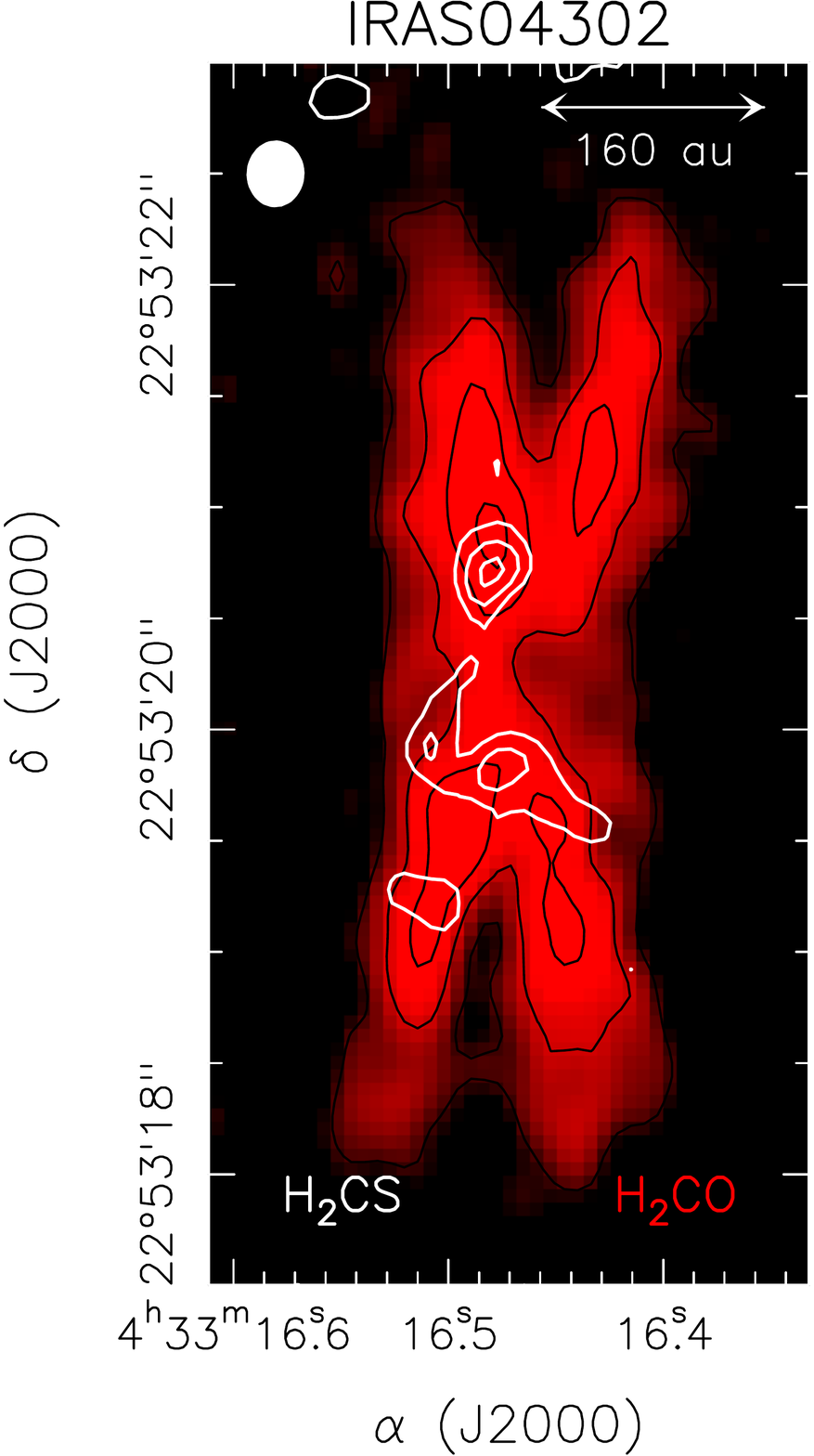}}
\caption{Spatial distribution (moment 0) maps (white contours) 
of the o--H$_2$CS(7$_{\rm 1,6}$--6$_{\rm 1,5}$) line (see Fig. 1), overlaid on the o-H$_2$CO($3_{1,2}-2_{1,1}$) reported by \citet{Podio2020}, in black contours and red scale.
Symbols are as in Figure 1.}
\label{comparison}
\end{figure} 

For the IRAS04302 disk, the H$_2$CS vertical intensity profile can be compared with those of CS and formaldehyde (H$_2$CO), by \citet{Podio2020}.
For such a purpose, we averaged the emission radially over 3 pixels (0$\farcs$18, corresponding to $\sim$ 29 au) around the selected radius.
Figure \ref{IRAS04302_radial} reports the comparison obtained at 60 au  and 115 au  from the protostar. 
Note that the H$_2$CS intensity 
has been multiplied by the factor reported in the labels in order to better
compare with those of o-H$_2$CO and CS.
The vertical distribution of H$_2$CS (black) shows an asymmetry with respect to the disk midplane (i.e. $z$ = 0), with the emission from the eastern disk side being brighter than in the western side by a factor 1.8 at 60 au radial distance and a factor 1.5 at 115 au. The same asymmetry in the vertical distribution is observed in the H$_2$CO (blue) and CS  (red) emission, and all the three species peak at a disk height, $z$ of about 50 au. This suggests that at our resolution the emission from H$_2$CS, H$_2$CO, and CS originate from the same disk layer.
As discussed by Podio et al. 2020, the bulk of the H$_2$CO and CS emission originate from the disk molecular layer where molecules are either released from dust grains and/or formed in gas-phase. Given that H$_2$CS  is co-spatial with H$_2$CO and CS and that all the three molecules can be efficiently formed in gas-phase using small hydrocarbons (e.g. CH$_3$), we suggest 
that H$_2$CS may form in similar way.

The comparison between the moment 0 distributions of H$_2$CS with those of H$_2$CO \citep{Podio2020} is also instructive (see Fig. 6), showing a slightly different radial distribution of H$_2$CO and H$_2$CS.}
The H$_2$CO (and CS) emissions extend radially out to $\sim 480$ au 
and peak at $\sim 120$ au, while the H$_2$CS emission is radially confined in the inner 115 au with a peak at $\sim 70$ au.
The radial distribution of H$_2$CS suggests that the o--H$_2$CS(7$_{\rm 1,6}$--6$_{\rm 1,5}$) line probes an inner portion of the IRAS04302 disk,
possibly due a higher $E_{\rm up}$, 60 K, with respect to those of the o--H$_2$CO(3$_{\rm 1,2}$--2$_{\rm 1,1}$) and CS(5--4), 33--35 K.
Only further observations of a species using different excitation lines will shed light on this hypothesis.

\section{Conclusions}

We presented the first images of the
radial (HL Tau) and vertical (IRAS04302) spatial distribution of the o-H$_2$CS emission as observed towards Class I protoplanetary disks using ALMA on a Solar System scale.
The observations have been performed in in the context of the ALMA chemical survey of Disk-Outflow sources in the Taurus star forming region (ALMA-DOT).
H$_2$CS is confined in a rotating ring with an inner dip towards the protostar: the outer radii are 140 au (HL Tau) and 115 au (IRAS04302). The edge-on geometry of IRAS 04302 allows us to reveal that H$_2$CS emission peaks, (at $r$ = 60--115 au), at $z$ = $\pm$ 50 au from the equatorial plane. Assuming LTE conditions, the column densities are $\sim$ 10$^{14}$ cm$^{-2}$. For HL Tau, we derived, for the first time, 
the [H$_2$CS]/[H] abundance in a protoplanetary disk ($\simeq$10$^{-14}$).
The o--H$_2$CS($7_{1,6}-6_{1,5}$) line emits where the emission of CS(5--4) and  o--H$_2$CO($3_{1,2}-2_{1,1}$) is brighter \citep{Podio2020}, i.e. the so-called warm molecular layer. 
H$_2$CS emission peaks closer to the protostar with respect to H$_2$CO and CS,
possibly due to the higher energy of the upper level (60 K) of the observed transition with respect to those of H$_2$CO and CS (33-35 K), which requires higher gas temperature hence favouring the emission from the inner disk regions.

The present work also provides 
the H$_2$CS and CS BEs as computed for the first time for
an extended crystalline ice (4258 K, H$_2$CS, and 3861 K, CS) 
and estimated for an AWS model using recipes from previous work \citep{Ferrero2020} to be in the 3000-4600 K and 2700-4000 K range. 
In turn, for $\sim$ 1 $L_{\rm \odot}$ protostars, this implies
that radially thermal desorption rules in an inner region,
while vertically only a thin upper layer is hot enough
\citep[see e.g.][]{Walsh2014,LeGal2019}.
To conclude, the observed H$_2$CS, more precisely
that detected at radii up to more than 100 au, is released into the gas likely due to non-thermal processes (photo-/CR- and/or reactive-desorption).

\begin{acknowledgements}
We thank the anonymous referee for instructive discussion and suggestions. This paper uses
the ADS/JAO.ALMA 2016.1.00846.S and
ADS/JAO.ALMA 2018.1.01037.S (PI L. Podio) ALMA data. ALMA is a partnership of ESO (representing
its member states), NSF (USA) and NINS (Japan), together with NRC
(Canada), MOST and ASIAA (Taiwan), and KASI (Republic of Korea), in cooperation with the Republic of Chile. 
This work was supported  
by the European Research Council 
(ERC) under the
European Union's Horizon 2020 research and innovation programmes:
(i) "The Dawn of Organic Chemistry" (DOC), grant agreement No
741002, and (ii) 
“Astro-Chemistry Origins” (ACO), Grant No 811312. CF acknowledges financial support from the French National Research Agency in the framework of the Investissements d’Avenir program (ANR-15-IDEX-02), through the funding of the "Origin of Life" project of the Univ. Grenoble-Alpes. DF acknowledges financial support from the Italian Ministry of Education, Universities and Research, project SIR (RBSI14ZRHR).
\end{acknowledgements}

\bibliographystyle{aa} 
\bibliography{./MasterReference.bib}

\end{document}